\documentclass[runningheads]{llncs}
\usepackage[T1]{fontenc}
\usepackage{graphicx}
\usepackage{hyperref}
\usepackage{color}
\hypersetup{
hidelinks,
colorlinks=true,     % false: boxed links; true: colored links
    linkcolor=blue,     % color of internal links
    citecolor=blue,     % color of citations
    filecolor=blue,     % color of file links
    urlcolor=blue        % color of external URLs
    }

\usepackage{blindtext} % Example package which populates the ever-popular lorem ipsum text.
\usepackage{listings}
\usepackage{xcolor}
\usepackage{url}
\usepackage{graphicx}
\usepackage{array}
\usepackage{makecell}
\definecolor{codegreen}{rgb}{0,0.6,0}
\definecolor{codegray}{rgb}{0.5,0.5,0.5}
\definecolor{codepurple}{rgb}{0.58,0,0.82}
\definecolor{backcolour}{rgb}{255, 255, 255}
\definecolor{codeblue}{rgb}{0,0,1}

\usepackage{listings} %code highlighter
\usepackage{color} %use color
\definecolor{mygreen}{rgb}{0,0.6,0}
\definecolor{mygray}{rgb}{0.5,0.5,0.5}
\definecolor{mymauve}{rgb}{0.58,0,0.82}
 
\definecolor{darkgray}{rgb}{.4,.4,.4}
\definecolor{purple}{rgb}{0.65, 0.12, 0.82}
 
\lstdefinelanguage{JavaScript}{
keywords={typeof, new, true, false, catch, function, return, null, catch, switch, var, if, in, while, do, else, case, break},
keywordstyle=\color{blue}\bfseries,
ndkeywords={class, export, boolean, throw, implements, import, this},
ndkeywordstyle=\color{darkgray}\bfseries,
identifierstyle=\color{black},
sensitive=false,
comment=[l]{//},
morecomment=[s]{/*}{*/},
commentstyle=\color{purple}\ttfamily,
stringstyle=\color{red}\ttfamily,
morestring=[b]',
morestring=[b]"
}
 
\begin{document}
\title{Exploring Bridges Between  Algorithmic and AI-generated Art}
\author{Jiaqi Wu\inst{1} \and
Eytan Adar\inst{1}}
\institute{University of Michigan, Ann Arbor MI 48109, USA
\email{\{wujiaq, eadar\}@umich.edu}}

\maketitle              % typeset the header of the contribution
\begin{abstract}
In this paper, we bridge algorithmic and AI art by adding functionality to the creative coding environment. We create two systems that demonstrate how AI features can enhance algorithmic art and, conversely, how AI art can be styled based on algorithmically-generated artifacts. The first library, \textit{GenP5}, extends p5.js to allow the artist to apply diffusion models to style and `condition' their algorithmically-constructed art. The second, \textit{P52Style}, can learn the `style' of an algorithmically generated artifact and apply that when creating new AI art. We provide all the code, demos, and art examples at \url{https://github.com/KolvacS-W/GenP5-P52Style}.

\end{abstract}
\section{Introduction}
\label{chpt:introduction}
Creative coding libraries and tools allow programmer-artists to create algorithmic art---expressive output crafted through code~\cite{Galanter2003WhatIG,peppler2005creative,shiffman2024nature}. Various programming languages and environments (e.g., Processing, p5.js, nodes~\footnote{\url{https://processing.org/}, \url{https://p5js.org/},\url{https://nodes.io/}}) are designed to make this type of creative coding efficient. Although powerful, algorithmic art environments have practical stylistic limitations. In this work, we demonstrate how new generative artificial intelligence (AI) technologies (e.g., diffusion models) can be combined with creative programming tools. This integration can expand the range of creative output and simplify complex styling tasks\footnote{Though `generative' art was traditionally used to describe versions of algorithmic, procedural, or software art, the term has since been co-opted (e.g., \textit{generative AI}). We will use \textit{AI art} to reference modern AI approaches such as Generative Adversarial Networks (GAN) or stable diffusion (SD) that are trained on data. We will use \textit{algorithmic art} when referring to art created through code, e.g., through p5.js.}.

Bridging these methods--the algorithmic and AI--represents a challenge and opportunity. There are many cases where creative programmers can make use of generative AI. For example, it is not easy for a p5 programmer to create artwork consisting of moving shapes in the style of Jackson Pollock. One solution is integrating the AI art approach into the algorithmic art environment. The artist could first generate the moving shapes programmatically and then apply a prompt-guided model (e.g., `angry acrylic splatters' with~\cite{meng2021sdedit}) to produce the desired look. Conversely, the artist can enhance their AI art by utilizing the `style' of algorithmically generated pieces. For example, the artist's algorithm produced colorful abstract tessellations. They could then apply the `style' to an AI-produced image of a city.

In this work, we explore how both kinds of tasks can be supported: where the AI generation can be applied to algorithmic art and where an algorithm's style can be applied to AI generation. We do both within the context of p5. Specifically, we contribute:

%In this work, we explore how to bridge creative coding and visual generative AI (specifically diffusion models) by programming functionalities integrated into the creative coding environment. Specifically, we want to explore methodologies to condition or stylize art content and perform style learning upon generative art via accessible interactions for artists and programmers. Specifically, we propose: 
\begin{itemize}
\item GenP5 (Section~\ref{chpt:GenP5}), a library enabling the styling and conditioned creation of algorithmic art using generative functions.
\item P52Style (Section~\ref{chpt:P52Style}), a library and tool that learns the algorithmic art's style and applies that when creating AI-generated art.
\item We reflect on how these two paradigms--programmed and AI-generated--can be integrated in future systems and libraries.
\end{itemize}

% #https://www.codecademy.com/learn/learn-p5js/modules/p5js-introduction-to-creative-coding/cheatsheet

\section{Background and Related Work}
\subsection{Algorithmic Art}
Algorithmic art (or sometimes procedural, software, generative, mathematical, creative code, etc.) is the process (and artifacts) of creative expression through the use of computer programming~\cite{peppler2005creative,shiffman2024nature}. It is generally considered the subset of generative art, which is the ``\textit{art practice where the artist uses a system, such as a set of natural language rules, a computer program, a machine, or other procedural invention, which is set into motion with some degree of autonomy contributing to or resulting in a completed work of art}~\cite{Galanter2003WhatIG}.'' These creative instructions can work in different ways. The artist Amy Goodchild holds that there are three main types of processes: randomness (random variables, noises, distributions, etc.), rules (algorithm instruction, mathematics formula, ecosystem simulation, etc.), and natural systems (e.g., growing biological system)~\cite{goodchild}. Though following the same principles, the specific manifestations of these artworks can vary from digital to physical, static to dynamic, and can be shown in all kinds of styles. By focusing on algorithmic art, we are interested in the generative art practice where artists ``\textit{program computers to undertake creative instructions}''~\cite{tempel2017generative}. Most often, algorithmic art is produced, ``autonomously or semi-autonomously,'' by a system~\cite{tempel2017generative}. The growing popularity of this community has expanded new media art practices~\cite{wiguna2022painting}. Our goal in this work is to create tools that can be used by such programmer-artists. However, we note that many of our approaches can be leveraged outside of this community.

In this work, we work within the p5.js `ecosystem.' P5.js is a Javascript branch of the popular Processing library/environment. The library allows programmers to create visually oriented applications with an emphasis on animation and interactions. It has been one of the most important tools for generative art creation and is popular among the creative coding and digital art communities. In addition to the core p5.js library, many users have contributed additional extensions. Our tools take a similar approach by working largely within this environment.

\subsection{Diffusion Models and its Applications}
% \blindtext[3]
Some of the most popular tools for AI art are based on Diffusion Models (DM) (e.g., Diffusion Probabilistic Model~\cite{ddpm}). Diffusion models learn to gradually `denoise' an image step by step (in reverse of their training) when guided by text, images, or other constraints. Such models are capable of generating images from pure noise given a text prompt. While the original diffusion models could be slow, advances such as the Latent Diffusion Model (LDM)~\cite{ldm} work with a lower-dimensional compressed representation of images to reduce memory and compute complexity. 

Other extensions (e.g., cross-attention mechanism~\cite{vaswani2017attention}), make it possible to use diffusion for other visual content-creation tasks. Tasks such as \textit{image stylization}, where one image can be stylized based on another, are effectively supported by these models (e.g., SDEdit~\cite{meng2021sdedit}). Instead of starting the diffusion process from the random latent image, the approach uses the input (reference) image as the initial latent by adding noise to it. Noise parameters control the influence of the initial image on the output. A similar task, \textit{style learning}, leverages the models' ability to learn the features of a customized visual style and use that style in the generation of new images. Popular models for style learning include a few-shot tuning~\cite{ruiz2022dreambooth}, textual inversion~\cite{gal2022image,ahn2023dreamstyler}, cross-attention~\cite{ye2023ip-adapter,wang2024instantstyle}, and Visual Style Prompting~\cite{jeong2024visual}. With GenP5 and P52Style, we build upon these models to support different combinations of AI and algorithmic art.

% focuses on self-attention layers of DMs, created a training-free method by swapping the self-attention of the target image inference guided by the inference of a reference image after inversion. For more details please refer to supplementary materials.

\subsection{Algorithmic Art and Generative AI}
Various artists and researchers have been exploring how to use generative AI in creative coding contexts. 
Liu el. al.~\cite{liu2023generative} used music audio as an input condition, deploying a large language model like GPT4 and a DM to generate music visualizations. SpellBurst~\cite{angert2023spellburst} is an authoring tool leveraging large language models to produce p5.js code. Various artists, including Takafumi Oyama, Roope Rainisto, and Brian Jordan~\footnote{see: \url{https://www.takafm.me/}, \url{https://linktr.ee/rainisto}, \url{https://bcjordan.com/}} used image or video models as a post-processing step to achieve creative effects. Dae In Chung and Brian Jordan~\footnote{see: \url{https://paperdove.com/about/} and \url{https://bcjordan.com/}} have been integrating generative AI into the programming stages (e.g., programming with conversational text and audio instructions).

Integrating diffusion models directly into a creative coding environment is still an under-explored area. In part, many algorithmic art projects are dynamic (e.g., abstract flowers growing from the ground). Many diffusion models were too slow or low resolution to keep up with the frame rate and detail of algorithmic art.
However, fast-inference approaches (e.g.,~\cite{luo2023latent}) have begun to enable real-time canvas stylization. As the algorithm renders each frame, it is sent to the diffusion model for stylization. For example, Dae In Chung created a tool to generate a stable diffusion image from any HTML5 canvas drawing~\footnote{\url{https://github.com/cdaein/vite-plugin-ssam-replicate}}. However, these approaches assume that the entire canvas should be styled the same way (e.g., as an oil painting). In many situations, an artist might want to stylize different components of their work independently or not at all (e.g., the background objects should be stylized as an oil painting, but the foreground objects should be unmodified). Our GenP5 library supports this kind of control by allowing for programmatic and independent control of the diffusion models.

\subsection{Conditions to Guide Algorithmic Art}
While algorithmic art often uses random noise, some artists prefer to use more structure. For example, an algorithmic art project might build a completely abstract artifact by laying fixed-size colorful lines on the screen at different angles. The effect, while having its own style, is more likely to be noise than to represent some structure. Instead, one could use an input image (e.g., a landscape photograph) to condition where lines are dropped and in what color. The piece, while still potentially abstract, can nonetheless be `read' as a landscape. Some research projects leverage these input conditions to create unique and creative artifacts that still satisfy some representational goal (e.g., text~\cite{batista2024evoboard} or portrait images~\cite{wieluch2024patternportrait}). As a specific example, Barile et. al.~\cite{barile2009animated} explored a way of generating animated images that start as noise but `resolve' to some underlying input picture. Based on this, Wu~\cite{wu2018saliency} designed a system to detect the input image's saliency map (e.g., the foreground/main subject of a photograph) and then apply paint strokes to gradually create a non-photorealistic image. Inspired by this approach, we added functions in GenP5 to allow artists to condition their algorithmic art.

%we would like to explore how to condition generative art with modular and encapsulated functionalities integrated into the creative programming language, to enhance the convenience of creating generative art with various conditions and improve the whole creative coding experience. This motivated us to extend the GenP5 library with functions to condition canvas contents.

\subsection{Style Learning}
In this work, we were also interested in how the style of algorithmic artwork could be applied in AI art. When using diffusion models, an artist often needs to be able to express their desired style through text or found images. In the case of algorithmic artists, each output of a program can serve as an example of the desired style. One inspiring example is Takafumi Oyama's \textit{Parametric Swimming}\footnote{\url{https://www.takafm.me/project/parametricswimming}}. In this work, the artist explored the connection between abstract art and real-life images by injecting the style of abstract art into real-world photos. However, since style learning methods require reference images as conditions, current approaches require manually adjusting and choosing the reference image as input to style diffusion models. As an alternative, we created P52Style. Our aim is to explore how we can augment algorithmic art style learning using diffusion models by integrating the whole process in a creative programming environment.

% Place your additional chapters here using the \input{} command
\section{GenP5}
\label{chpt:GenP5}
We describe GenP5, a library for p5.js to support the use of diffusion models in creating algorithmic art. GenP5 provides (1) the ability to apply diffusion-based stylization to specific visual elements and (2) the use of diffusion output to condition the algorithmic process (e.g., having generated particles create a heart shape).

\begin{figure}[t]
    \centering    \includegraphics[width = \linewidth]{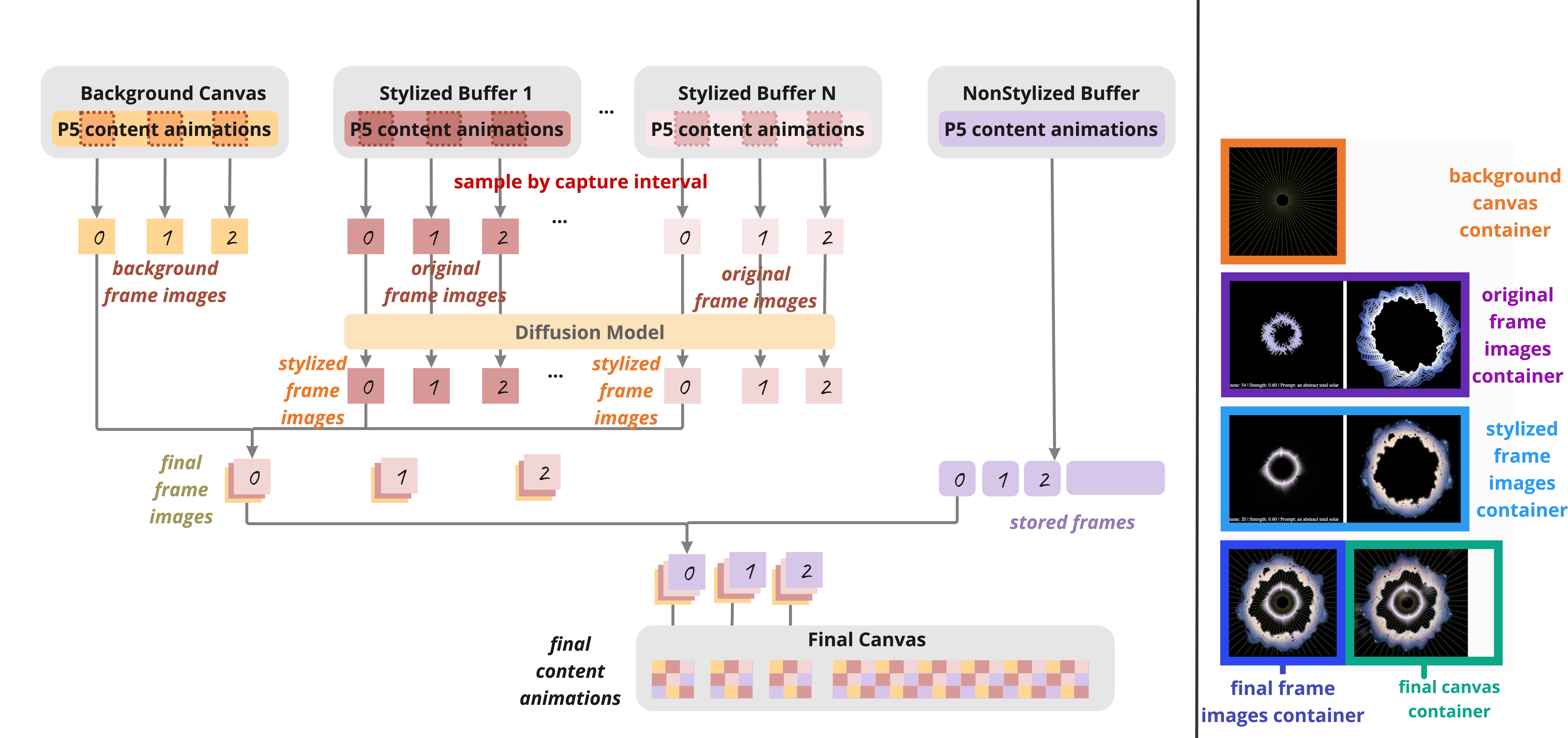}
    \caption{Left: GenP5 method overview. Numbers indicate a \textit{frame index}. Right: UI elements that will be dynamically created when GenP5 library is used. In this example, a slight background pattern is drawn in the \textit{background canvas}. Two \textit{stylized buffers} are created, each containing a ring-shape animation. One \textit{nonstylized buffer} contains bubble effect filters (this buffer is not displayed in the UI).}
    \label{fig:genp5abstract}
\end{figure}

\subsection{Method}

\subsubsection{Stylizing Canvas Contents}
Figure~\ref{fig:genp5abstract} represents GenP5's architecture or stylization. At the highest level, the output of the p5.js can be selectively stylized using a diffusion model for each frame (i.e., each step of algorithmic production). The library is predominantly based on buffers and canvases.

The \textit{background canvas} is the p5.js drawing surface that is created by default for any p5.js project (one draws onto this surface). To stylize all or part of rendered objects, GenP5 uses \textit{stylized buffers}. These are off-screen graphics into which the code can draw. Anything within these buffers will be stylized using a diffusion model (guided by a textual prompt). \textit{Nonstylized buffers} can also be created and drawn to, but these will not be passed to a diffusion model. The background canvas and off-screen buffers can be combined to produce a final frame for display. As a specific example, we might render a bouncing circle into a stylized buffer (with the prompt `ball as oil painting'). We also render a rotating square into a nonstylized buffer in front of it. The combined animation will show a bouncing oil painting ball behind a rotating (vector art) square.

A frame (with an associated index) consists of a set of background canvas, stylized and nonstylized buffers (each with the same index). These correspond to the animation frame that p5.js would ordinarily render to the screen. As stylized buffers are captured, their content is passed to a diffusion model to generate an image. Buffers (stylized and non) are stacked to produce a final image. To ensure that buffers do not obstruct each other when combined, we remove the background from these images~\footnote{The algorithm identifies the most frequent color in the image as the background color, iterates through image pixels, setting the pixel's alpha channel is set to 0 when it considered close to the background color.}. For nonstylized buffers, GenP5 uses the properties of the p5.js objects rendered to that buffer (e.g., transparency).

When using the GenP5 library, we automatically create a set of additional UI components. These can be used to debug the frames as the program is run. As shown in Figure~\ref{fig:genp5abstract} (right), the \textit{background canvas container} contains the \textit{background canvas}. Other components show the \textit{original frame images}--the content of each of the stylized buffers \textit{before} they are passed to the diffusion model. Separate containers will show the stylized buffer images \textit{after} diffusion has been applied. One \textit{final image container} shows the combined output. The final output update at the framerate specified in the program.

\begin{figure}[htbp]
    \centering    \includegraphics[width = \linewidth]{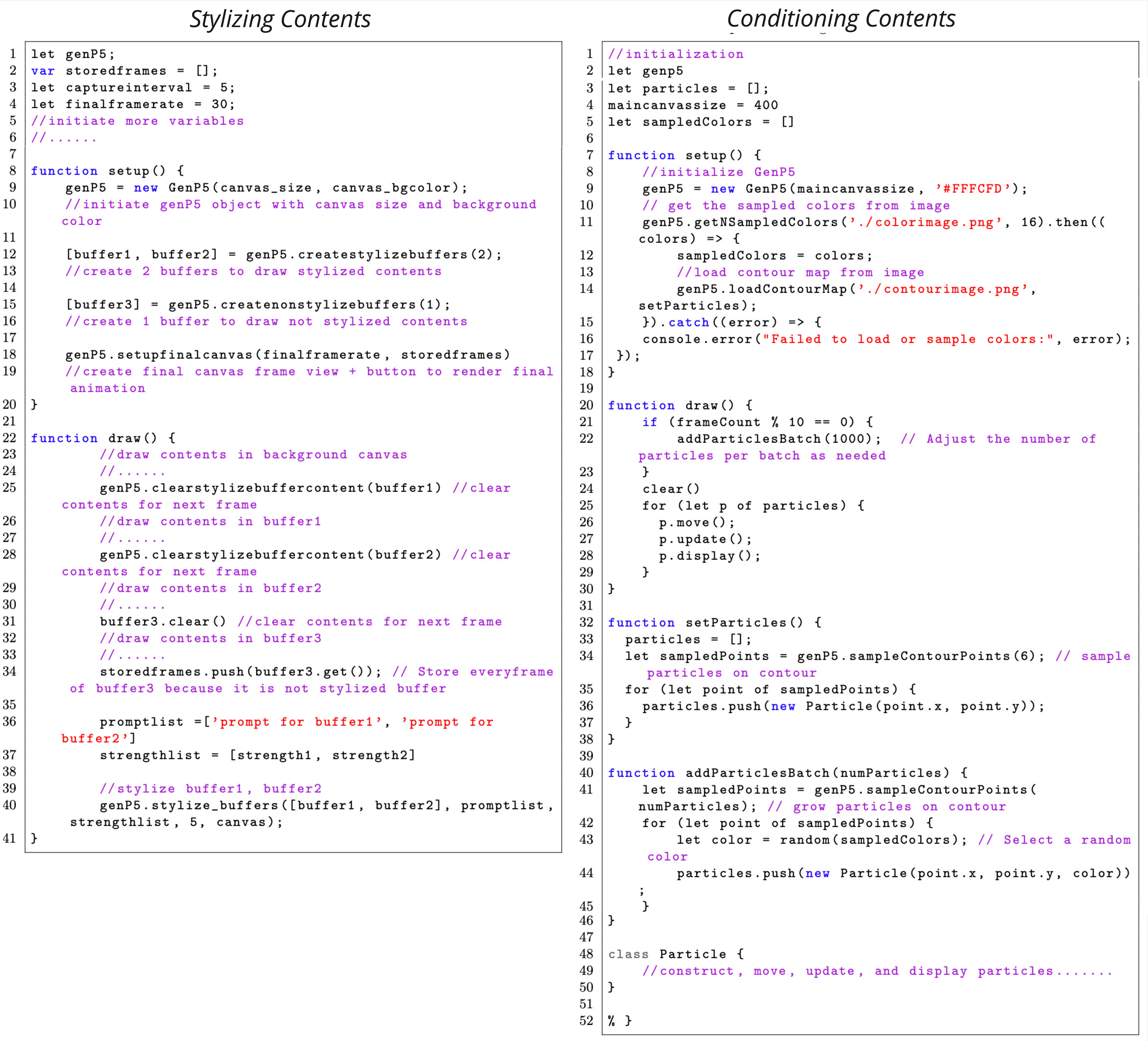}
    \caption{Code snippet of creating art projects with GenP5 library.}
    \label{fig:code}
\end{figure}

Figure~\ref{fig:code} (left) illustrates a simple p5.js program with content stylization. The API was designed to require minimal adaptation to existing p5.js programs. For example, simple functions allow the programmer to create stylized buffers. Once these are created, drawing into these works in the same way as a standard p5.js program. Complete details of the API are available in the supplement.

\begin{figure*}[ht!] 
     \centering\includegraphics[width=1\linewidth]{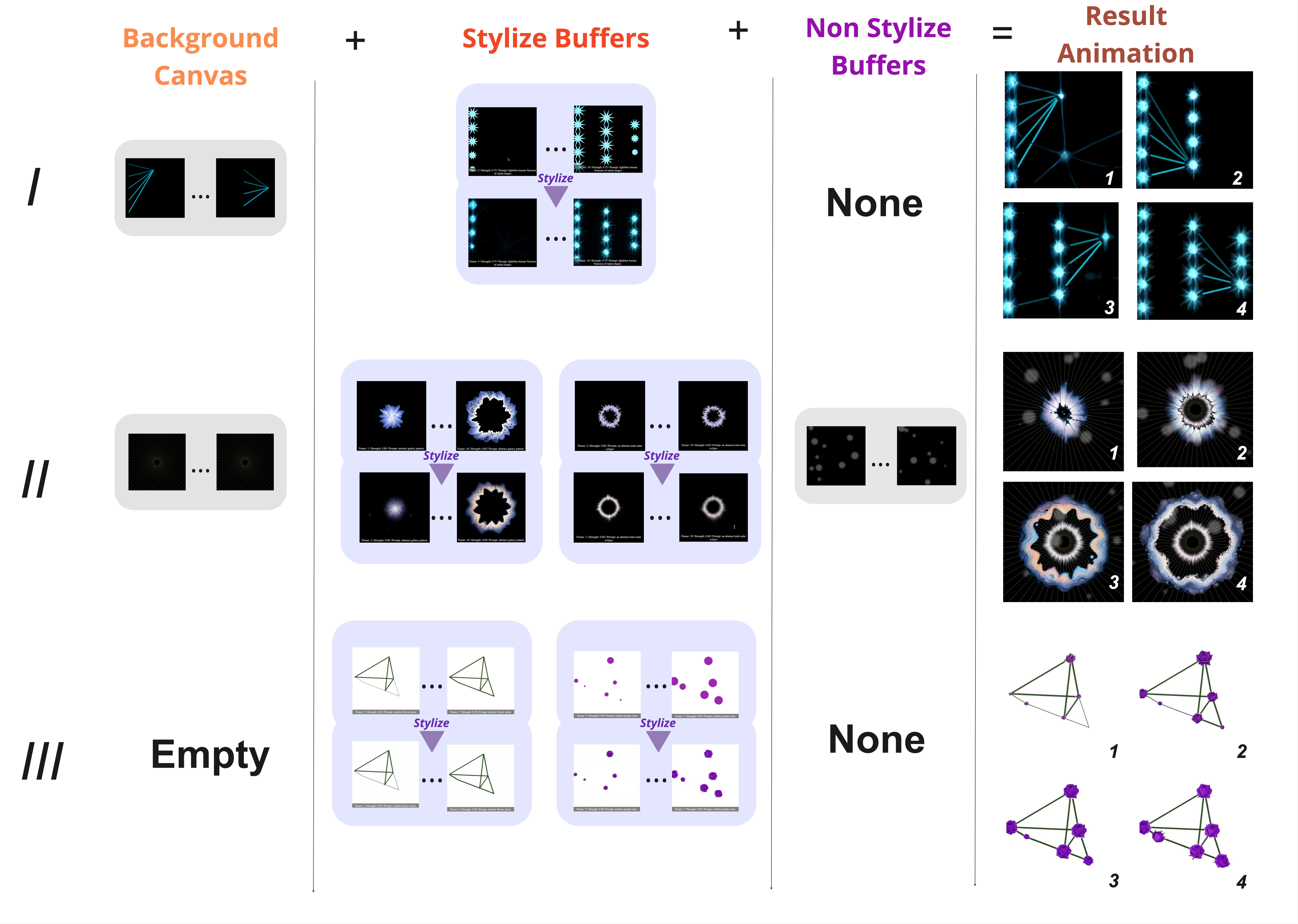}
    \caption{Three simple examples of generative art
project using GenP5 functions for stylization. Prompts for stylize buffers: `lightblue human neurons of radial shapes' (row I), `abstract galaxy pattern' (row II left), `an abstract total solar eclipse' (row II right), `realistic flower stems' (row III left), `realistic purple roses' (row III right).  }
    \label{fig:genp5examplenew}
\end{figure*}

Figure~\ref{fig:genp5examplenew}, provides three examples of GenP5's stylization library. The first row (line I) is a stylized neural network visualization. The program stylizes the (neural) nodes while keeping the content that needs to be structurally precise (the edges) non-stylized in the background. The second example (line II) is an abstract animation inspired by Eclipse. Non-stylized content is placed in a separate layer to preserve effects that cannot be shown by generated images. Finally, the third row depicts frames from an animated simulation of a graph search algorithm. Nodes and edges are placed in separate stylized buffers (with flowers representing nodes and stems as edges). Here, the stylized buffers are independently controlled with a `strength' parameter that can help control how much of the input drawing is retained.

\subsubsection{Conditioning Canvas Contents}
In addition to stylization, GenP5 also supports conditioning. The artist can supply a textual specification describing the pattern they would like to use in their algorithmic art (e.g., colors, shapes, etc.). For example, they can specify that rather than randomly moving, the p5.js particles should follow a heart pattern. Static images can be used for this, but patterns can be described through a textual prompt that is fed to the diffusion model. This process introduces its own kind of randomness as each run of the diffusion model can output a different object (e.g., different heart shapes or different colors extracted from a diffused image of a sunset). We view this kind of variability as desirable as much of algorithmic art produces randomized output every time the program is executed. In the current instantiation of the library, we provide mechanisms for extracting the contours of the object in the generated image (i.e., the outline of the heart) as well as sampling colors. However, programmers have access to the diffusion-generated images and can extract other features for use in their art.

\begin{figure*}[ht!] 
     \centering\includegraphics[width=0.8\linewidth]{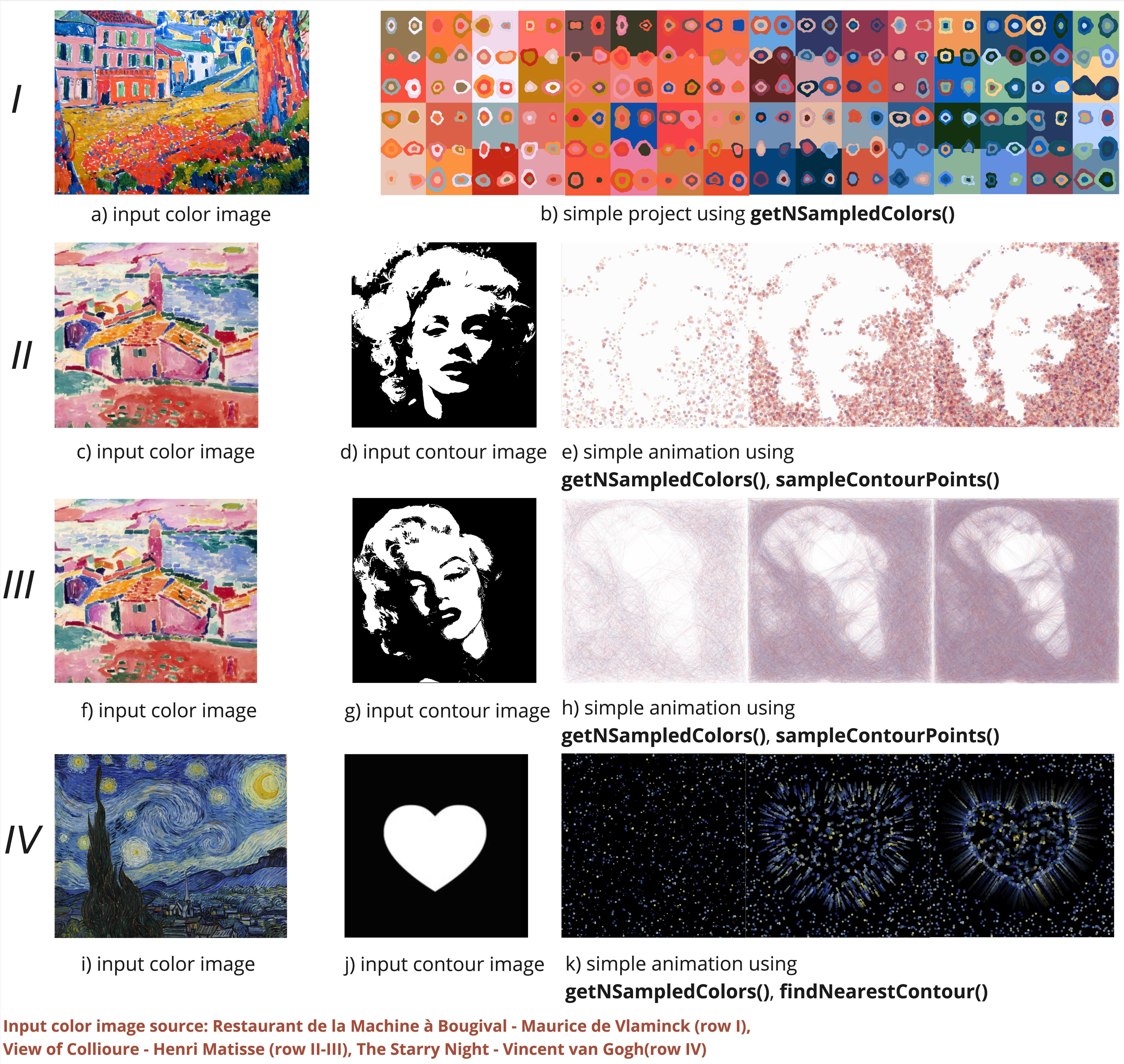}
    \caption{Four examples of generative art
project using GenP5 functions for conditioning. The input contour maps are inverted.}
    \label{fig:conditionexample}
\end{figure*}

Figure~\ref{fig:code} (right) demonstrates the use of the conditioning feature (see example II in Figure~\ref{fig:conditionexample}). Starting in line 11, the program extracts the colors (\textit{getNSampledColors()}) and contour map (\textit{loadContourMap()}) of the image (these images were previously created and saved to simplify the example). The moving particles in this image will `stick' to the randomly sampled points in the contour map (\textit{sampleContourPoints()}) and will randomly use the colors extracted from the image. Other functions allow the programmer to find the nearest contour point  (\textit{findNearestContour()}), allowing the particles to `snap' to the closest point.

Figure 4 illustrates four examples of conditioning. These represent the variety of combinations that can be created. Note that in these specific examples, we use `real' (static) images rather than diffusion-generated examples to make the effect clear.

\subsection{Usage and Implementation Detail}
In our prototype implementation, the GenP5 utilizes an external server to run diffusion models. Specifically, a web socket is created between the client library and the locally hosted server (implemented as a node.js project). To ensure that diffusion can happen in real-time regardless of the web browser or server's capabilities, we leverage the fal.ai LCM model~\footnote{see \url{http://fal.ai}}. Fal.ai charges a nominal amount for running the diffusion model, but GenP5 can be modified to use models hosted elsewhere (even locally). Our server handles queuing and saving stylized frames.

\section{P52Style}
\label{chpt:P52Style}
The GenP5 library provides artists with a way to apply features of AI art to their algorithmic artifacts. In this section, we describe a second library, P52Style, that supports the application of algorithmic artifacts to AI art. Specifically, the system captures the `style' of the algorithmically generated art and uses these to guide the diffusion model when generating content. P52Style provides a library and GUI (built on p5.gui\footnote{\url{https://bitcraftlab.github.io/p5.gui/}}) to control the algorithmic output and a set of functions to extract stylistic aspects of this output.

In this paper, we define \textbf{style learning task} as generating new images with a customized style `learned' from reference images that have the target style. In standard AI art, the artist must identify sufficient reference images from which the style can be extracted. This may be difficult if these images do not exist or are not varied enough to capture the style in enough detail. Additionally, it is possible that there is no existing example of the style, nor can it be explained in natural language. In this case, the artist may want to \textit{program} examples of the style. P52Style supports these workflows.

Style is often difficult to capture in one image. Algorithmic art, specifically variants that use randomization, can produce multiple reference images. However, many algorithmic art projects have many parameters, and waiting for output that captures the desired style may be tedious. Rather than relying on pure randomness to (hopefully) generate good reference images, P52Style adds GUI elements to more precisely control the generated output. This allows the artist to control, in real time, the content being rendered. In addition to adding a GUI dashboard for control, P52Style also captures all rendered frames. The artist can rapidly move back and forth in the rendered sequence to find frames that will work as reference images.

\begin{figure}[ht]
    % \centering
    \includegraphics[width = \textwidth]
    {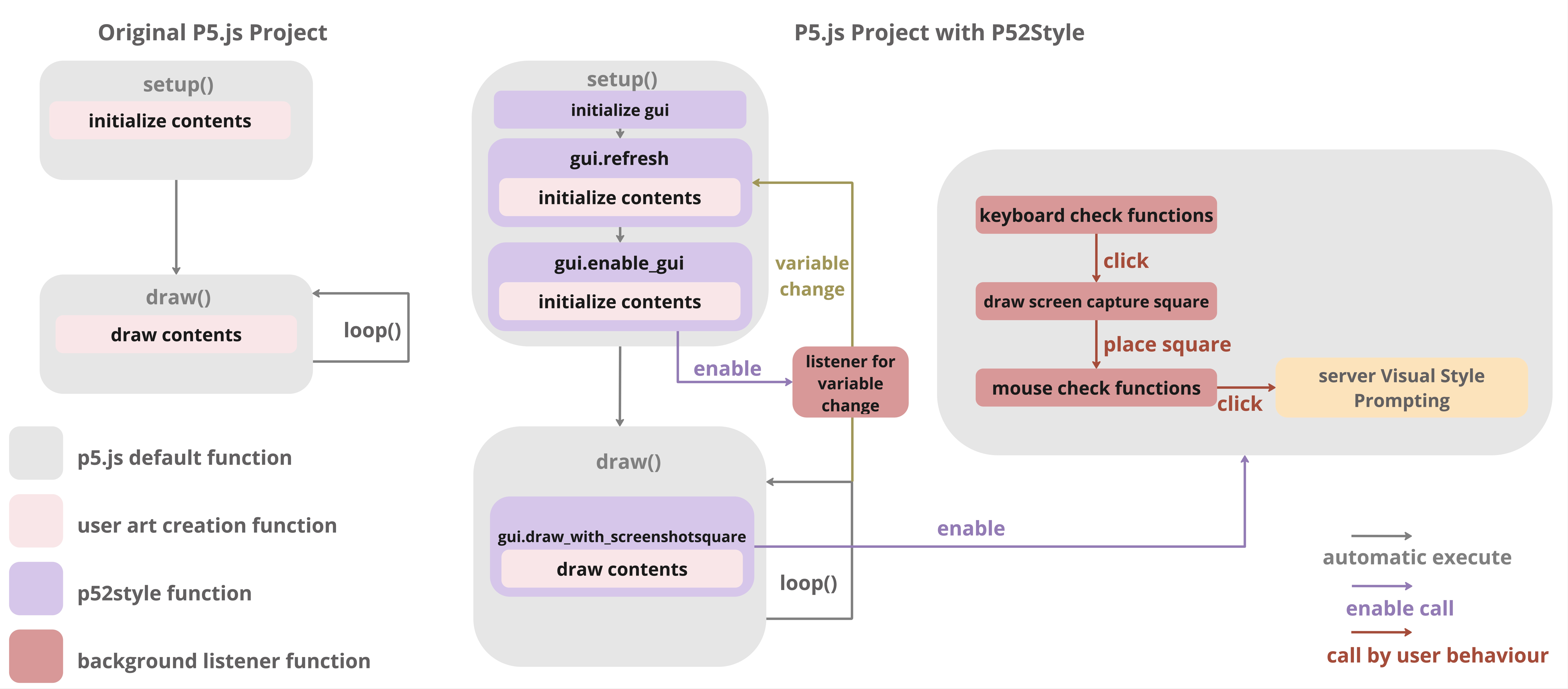}
    \caption{Overview of p52style structure.}
    \label{fig:p52style_structure}
\end{figure}

\subsection{Method}

\paragraph{Overview}
As shown in Figure~\ref{fig:p52style_structure}, we view all p5.js project as having three parts: functions to initialize content, functions to draw the content, and other supporting functions (eliminated in the figure). To use a p5.js project in P52Style, we wrap up the function to initialize content and the function to draw content separately as callback functions for our library. Specifically, we provide function calls to: 
\begin{enumerate}
\item Add all the variables as GUI components \textit{ (initialize\_gui())} 

\item Restart the whole animation \textit{(refresh())}

\item Enable the listeners to restart the whole animation on change of any variables \textit{(enable\_gui())}

\item Enable the frame index selector and screenshot capture logic \\ \textit{(draw\_with\_screenshotsquare())}
\end{enumerate}

\begin{figure}[htbp]
    \centering
    \includegraphics[width = 0.8\textwidth]
    {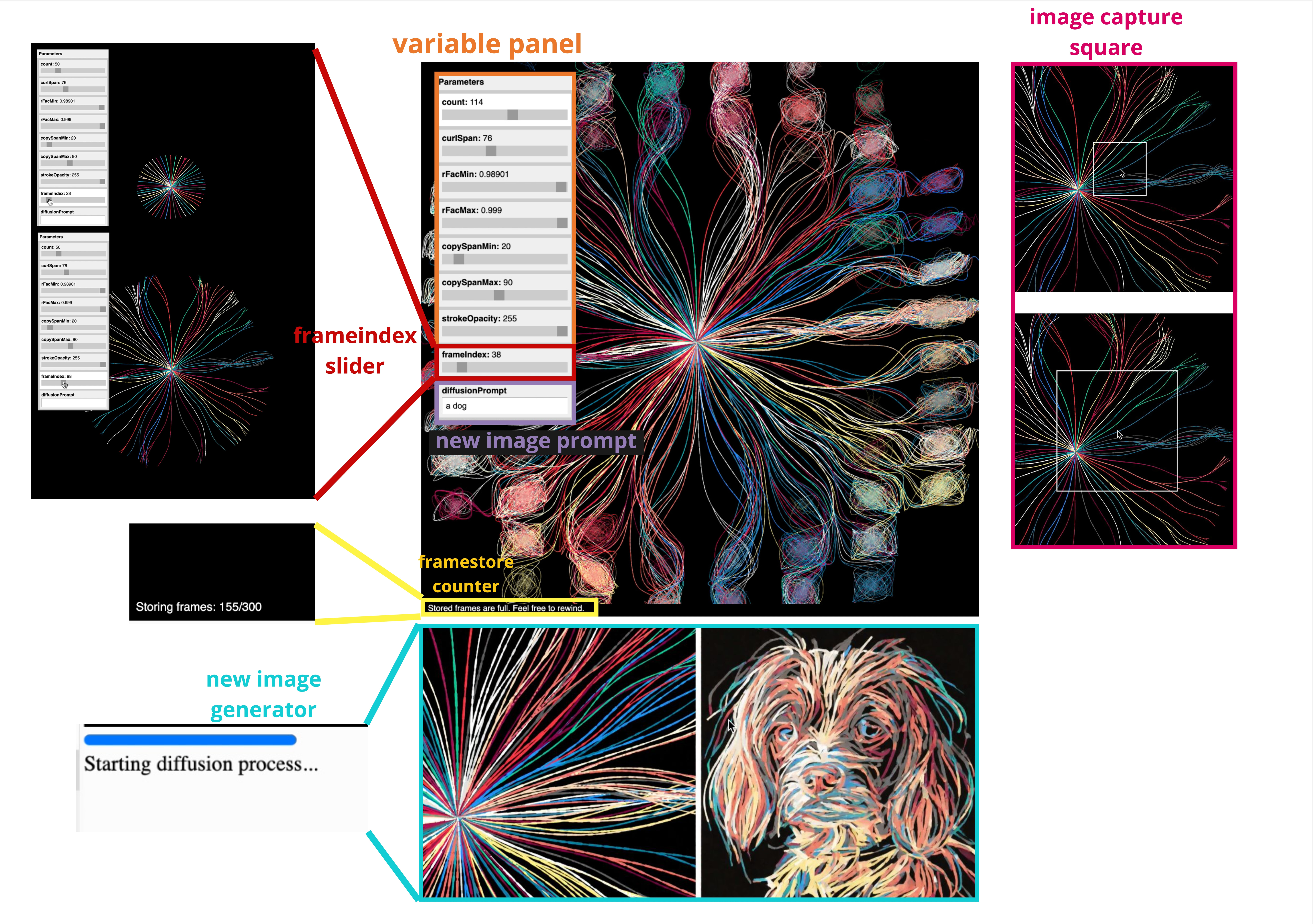}
    \caption{Overview of p52style UI. Example p5 art credit:  Inner Demons by Che-Yu Wu}
    \label{fig:p52style_ui}
\end{figure}

\paragraph{P52Style GUI}
When users enable the P52Style library in p5.js, we automatically generate UI components. As shown in Figure~\ref{fig:p52style_ui}, the core elements of the UI are:
\begin{enumerate}
\item Variable panel: Built on top of the p5.gui panel, this contains all the variables of the art. Any change of the variable will refresh the whole art animation.

\item Framestore counter: Every time the art is refreshed, the counter will automatically store $N$ frames and show the storing progress. $N$ is specified by the user in function calls.

\item Frameindex slider: Once the framestore counter shows the completion of storing frames, the slider allows the user to adjust and rewind to any frame number (1--$N$), and conveniently select any stage of the whole animation.

\item New image prompt: Allows the user to specify the prompt to generate new images for style-learning tasks.

\item Image capture square: when the user presses “Tab” on the keyboard, an image capture square will be evoked, allowing the user to select any image patch to use as a reference style image for style learning. Users can adjust the position and size of the square with a keyboard and mouse.

\item New image generator: After the user clicks on the ideal location of the capture square, the system will call Visual Style Prompting with the new image prompt and captured style reference image, showing a new image generation process below the canvas. After the image is generated, it will automatically be shown below the canvas and downloaded locally.
\end{enumerate}

\subsection{Implementation Details}
We investigated\footnote{For results and additional details, please refer to supplementary materials} the state-of-the-art style learning methods described in Section~\ref{chpt:introduction}. We found that Visual Style Prompting~\cite{jeong2024visual} was the best method to capture the style units and generate an image corresponding to the prompt description. On the client side, P52Style is implemented by extending the p5.gui library. As Visual Style Prompting is best done on a sufficiently powered machine with a GPU, we built a server built into the Google Colab notebook (using an A100 GPU). The frontend library connects to this server to perform the style prompting operations.

\section{Usability Evaluation}
As there are no equivalent libraries to GenP5 and P52Style, we provide a simple evaluation using Cognitive Dimensions of Notations (CDs)~\cite{blackwell2001cognitive}. We score each dimension as high, low, mid, and standard (indicating the same level as any other p5.js or programming libraries, the level is acceptable as long as the user has proper knowledge of p5.js programming). We describe why we score each dimension, but these scores can be enhanced through observation of use. We leave a more formal user study to future work.

\subsection{Low-scoring Dimensions}
\textit{Viscosity: Resistance to change}. The library is not viscous, and the changes in content code do not require changes in function calls. The changes can be easily made by adjusting function parameters or adding/deleting function calls.

\vspace{3pt}
\noindent
\textit{Secondary Notation: Extra information in means other than formal syntax}. 
The library functions do not require users to record extra information like comments.

\vspace{3pt}
\noindent
\textit{Diffuseness: Verbosity of language}. 
The function calls have low diffuseness, as a feature of programming languages. For UI components, the label text is reduced to the minimum.

\subsection{Mid-scoring Dimensions}
\textit{Visibility: Ability to View Components Easily}. The library has simple UI components with minimum/no labels and captions.

\vspace{3pt}
\noindent
\textit{Hidden Dependencies: The way the user uses one library function will not affect other functions implicitly}. Certain user code behaviors concerning the canvas display when using P52Style might damage the performance of the library (e.g., the \textit{blend()} call).

\vspace{3pt}
\noindent
\textit{Error-Proneness: The notation invites mistakes and the system gives little protection.} 
The library functions have simple and explicit usage, but there might be user mistakes with no clear system feedback.

\subsection{Standard-scoring Dimensions}
\textit{Premature Commitment: Constraints on the Order of Doing Things}. Like any other p5.js library, initiating instances, including package code, and a certain sequence of executing is necessary.

\vspace{3pt}
\noindent
\textit{Abstraction: Types and availability of abstraction mechanisms}.
The library functions have the same abstraction level as other p5.js libraries.

\noindent
\textit{Hard Mental Operations: High demand on cognitive resources}. 
The function calls have a standard level of hardness as programming languages. For UI components, the icons and text are easy to understand without cognitive burden.

\vspace{3pt}
\noindent
\textit{Provisionality: Degree of commitment to actions or marks.} 
The library has a standard level of provisionality as any other programming library.

\noindent
\subsection{High-scoring Dimensions}
\textit{Role-Expressiveness: The purpose of an entity is readily inferred}. 
The library follows the cognitive habit of users as it has the same structure as any other p5.js library. The purpose of different functions is well elaborated with function names and documentation.

\vspace{3pt}
\noindent
\textit{Closeness of Mapping: Closeness of representation to domain}. 
The purpose of different functions is well elaborated with function names and documentation, whereas the closeness of functions and behavior is standard in programming libraries. 

\vspace{3pt}
\noindent
\textit{Consistency: Similar semantics are expressed in similar syntactic forms}. 
The library has high consistency, as a feature of programming languages.

\vspace{3pt}
\noindent
\textit{Progressive Evaluation: Work-to-date can be checked at any time.} 
The library has a high progressive evaluation as any other programming library.

\section{Conclusions and Future Work}
In reviewing how p5.js projects might use GenP5, we identified situations where more control of the (un)stylized content is necessary. Future extensions of the library can provide additional control over how and when buffers are passed to the diffusion model. Additional masking functions may also provide better ways to stack the layers. While the prototype utilizes a client/server model for executing the diffusion, new models may be run in the browser and provide more performant and high-resolution output. The first version of GenP5 provides examples of conditioning for contours and colors. We hope to work with our users to identify other possibilities and use cases and to add these functions.

For P52Style, we would like to extend the library to support flexible manipulation of a large number of variable types, including color selection and random variables. In our experience, we have noticed that it is sometimes difficult to determine which variables need to be changed (and by how much) to achieve interesting variations in the output images. We are considering better displays that can pre-calculate examples to better guide the artist in finding `good' settings. For both GenP5 and P52Style, we intended to conduct additional controlled experiments as well as release the software to get feedback from external users.

In this work, we investigated how algorithmic and AI art can be bridged in the context of existing creative coding environments. We introduce GenP5 and P52Style as tools for the creation of artifacts that utilize both algorithmic and AI approaches synergistically. Our initial analysis and case studies demonstrate the viability of the approach and the possibility for new forms of creative expression.

\bibliographystyle{splncs04}
\bibliography{references}

\end{document}